\documentclass[aps]{revtex4}
\usepackage{amsmath}
\usepackage{graphicx}
\usepackage{hyperref}

\newcommand{\snn}{\sqrt{s_{\rm _{NN}}}}
\newcommand{\tauf}{\tau_{\rm _F}}

\begin{document}

\title{A Semi-analytical Method of Calculating Nuclear Collision
  Trajectory in the QCD Phase Diagram}
\author{Zi-Wei Lin}
\affiliation{Department of Physics, East Carolina University,
  Greenville, NC 27858}
\author{Todd Mendenhall}
\affiliation{Department of Physics, East Carolina University,
  Greenville, NC 27858}

\begin{abstract}
The finite nuclear thickness affects the energy density $\epsilon(t)$
and conserved-charge densities such as the net-baryon density $n_B(t)$
produced in heavy ion collisions. While the effect is small at
high collision energies where the Bjorken energy density formula 
for the initial state  is valid, the effect is large at low collision
energies, where the  nuclear crossing time is not small compared to the
parton formation time. The temperature $T(t)$ and 
chemical potentials $\mu(t)$ of the dense matter can be extracted from the
densities for a given equation of state (EOS). Therefore, including the
nuclear thickness is essential for the determination of the
$T$-$\mu_B$ trajectory in the QCD phase diagram for relativistic
nuclear collisions at low to moderate energies such as the RHIC-BES
energies. In this proceeding, we will first discuss our
semi-analytical method that includes the nuclear thickness effect and
its results on the densities $\epsilon(t), n_B(t), n_Q(t)$, and
$n_S(t)$. Then, we will show the extracted $T(t), \mu_B(t), \mu_Q(t)$,
and $\mu_S(t)$ for a quark-gluon plasma using the ideal gas EOS with
quantum or Boltzmann statistics. Finally, we will show the results on
the $T$-$\mu_B$ trajectories in relation to the possible location of
the QCD critical end point. 
This semi-analytical model provides a convenient tool for exploring
the trajectories of nuclear collisions in the QCD phase diagram. 
\end{abstract}

\maketitle

\section{Introduction}

The study of the QCD phase diagram, including the possible 
critical end point (CEP) that separates the crossover transition
from a first-order transition, is a focus of relativistic heavy
ion physics\cite{Stephanov2005,Aggarwal2010,STARCollaboration2020}.  
For this purpose, it is important to calculate or estimate the
collision trajectory  in the QCD phase diagram in the $T-\mu_B$ plane
or the general  $T-\mu_B-\mu_Q-\mu_S$ four-dimensional space.  
So far, this is mostly done by analyzing the time evolution of
given volume cells in dynamical
models~\cite{Arsene:2006vf,Wang:2021owa} such as transport models 
and hydrodynamical models, which usually takes a lot of effort. 
A trajectory can also be estimated by using a given
equation  of state with assumptions such as imposing a constant
$s/n_B$ (entropy to net-baryon-density ratio) 
along the trajectory~\cite{Noronha-Hostler:2019ayj}. In this approach, 
extra information is needed to determine the endpoint of the
trajectory at the maximum energy density. 

Here we present a semi-analytical method~\cite{Mendenhall:2021maf} to
calculate the trajectory of central nuclear collisions in the QCD
phase diagram. It is straightforward to reproduce and already
available online~\cite{interface}. It also allows us to have
analytical understanding of the effects of the finite nuclear
thickness and parton formation time on the collision trajectories of
different collision systems at different energies.  
The method first calculates the time evolution of four densities: 
the energy density $\epsilon$, net-baryon density $n_B$,
net-electric-charge density $n_Q$, and net-strangeness density $n_S$. 
We then use a given EOS to convert them to four thermodynamic
quantities: $T,\mu_B,\mu_Q$, and $\mu_S$.

A famous semi-analytical result on the energy density 
is the Bjorken energy density formula for
mid-spacetime-rapidity~\cite{Bjorken1983}: 
\begin{equation}
\epsilon^{Bj}(t)=\frac{1}{tA_T}\frac{dE_T}{dy}.
\label{e_bj}
\end{equation}
In the above, $A_T=\pi R_A^2$ is the transverse overlap area of the
two nuclei in central collisions, $R_A=1.12A^{1/3}$ fm is the nuclear radius in the
hard-sphere model,  and $dE_T/dy$ is the transverse energy rapidity density at
mid-rapidity.  The formula is mostly used to estimate the energy
density of the initial state, where time $t$ is chosen as the
formation time of the quark-gluon plasma or the produced partons
($\tau_F$).  Note that the formula also applies to any time $t$ after
$\tau_F$ when the partons are assumed to be free-streaming; it thus
describes the time evolution of  the energy density produced from the
initial state when the subsequent parton interactions and transverse
expansion are neglected.

\begin{figure}[hbt]
   \begin{minipage}[t]{0.6\linewidth}
   \centering
   \includegraphics[width=0.95\linewidth]{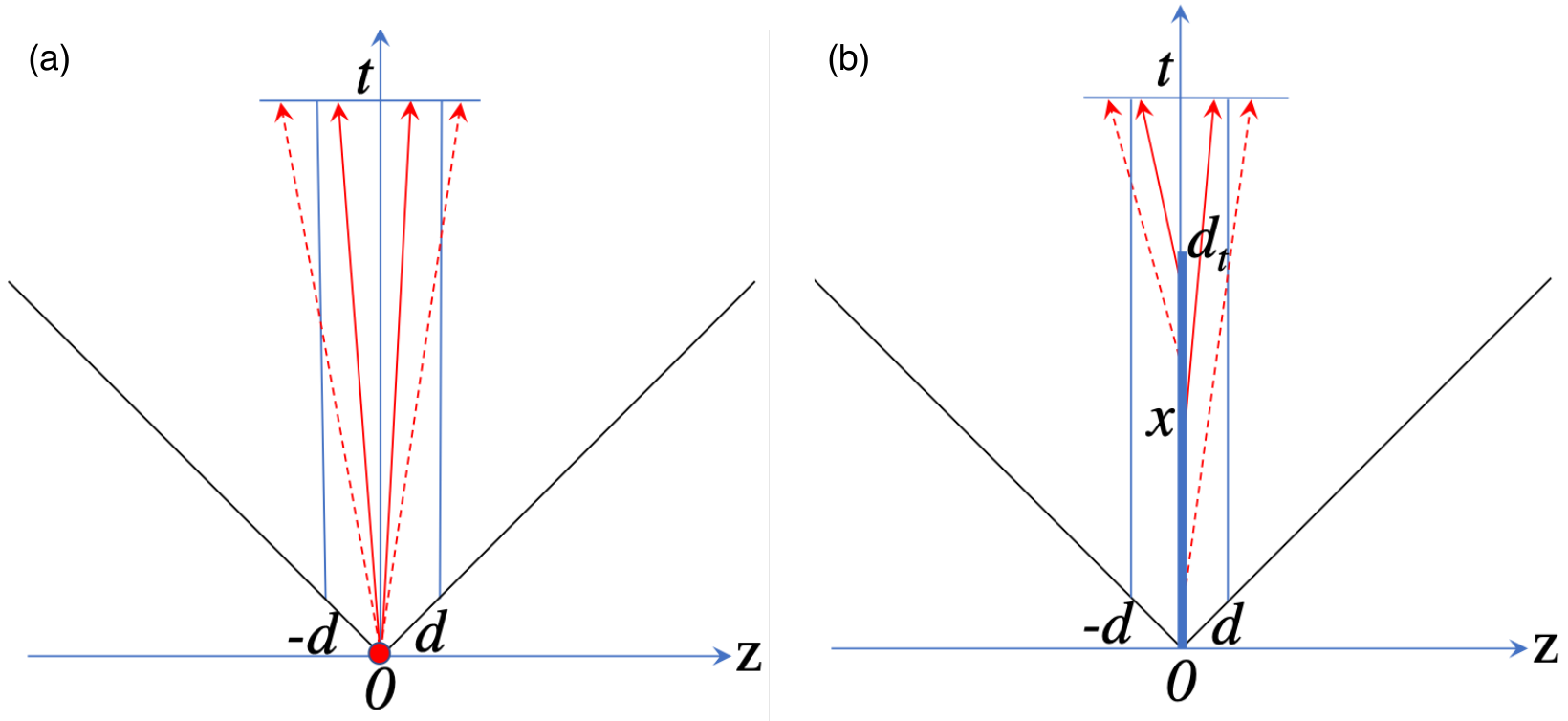}
   \end{minipage}%
   \begin{minipage}[t]{0.38\linewidth}
   \centering
   \includegraphics[width=0.95\linewidth]{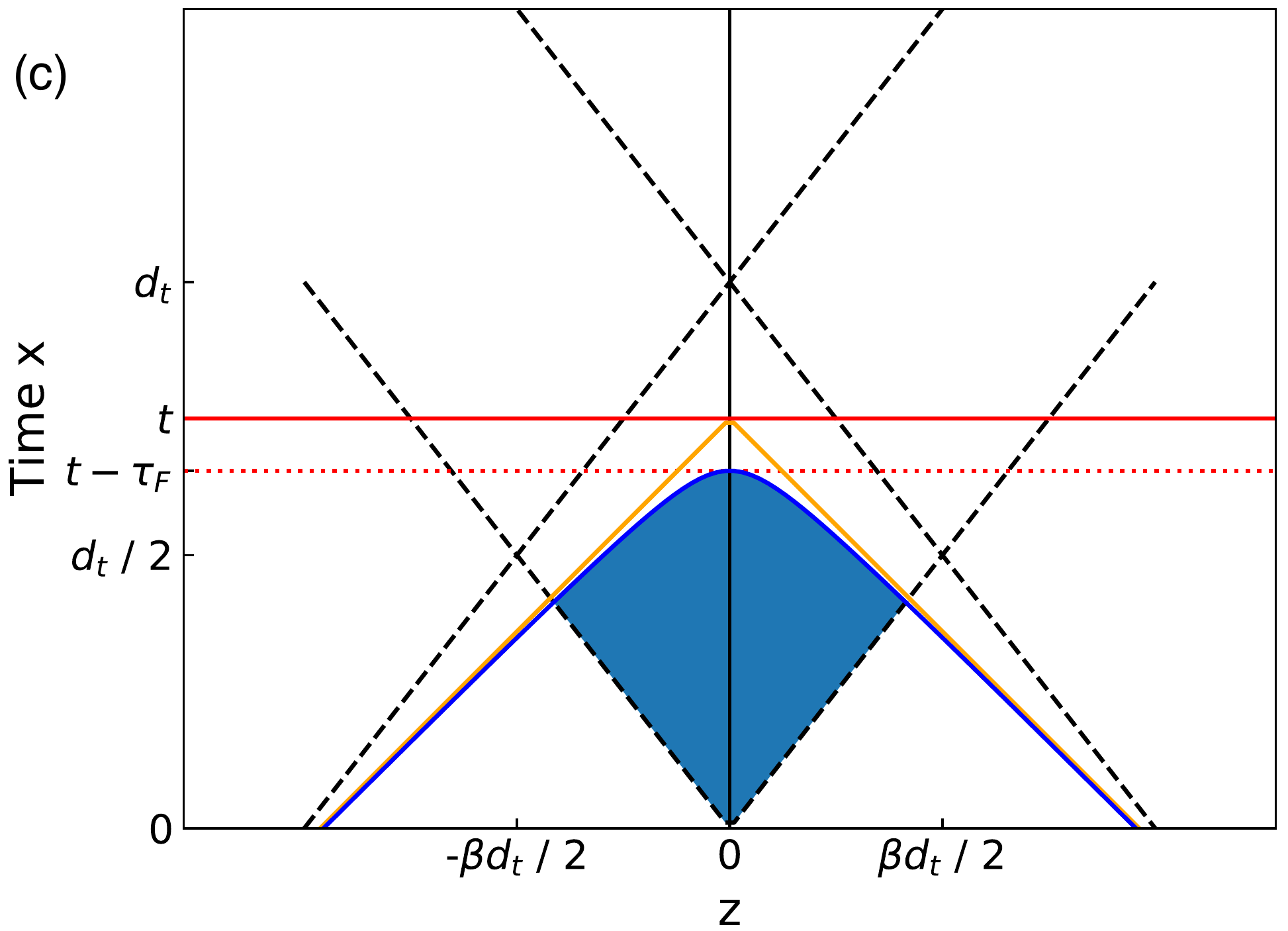}
   \end{minipage}
\caption{Schematic diagram for the crossing of two identical
nuclei for (a) the Bjorken $\epsilon$ formula, 
(b) the method that considers the finite crossing time but not the
finite longitudinal width \cite{Lin2018}, and (c) the
current method that considers the full crossing diamond area
\cite{Mendenhall2021}. 
In (c), partons can be produced anywhere inside the
rhombus, the solid diagonal lines represent the light cone
boundaries for partons that can reach $z   \approx 0$ at time $t$,
while the hyperbola represents the   boundary of these partons after
considering the formation time $t_F=\tau_F\,{\rm \cosh}\,y$.}
\label{fig1}
\end{figure}

However, the Bjorken energy density formula breaks down at low energies
where the nuclear crossing time is not small compared to $\tauf$. 
In the hard sphere model for the nucleus, the crossing time is given 
by $d_t=2R_A/(\beta \gamma)$, where $\beta$ is the speed of the 
projectile nucleus in the center-of-mass frame and
$\gamma=1/\sqrt {1-\beta^2}$ is the corresponding Lorentz factor. 
Obviously, the crossing time becomes bigger at lower energies, 
where its effect must be considered.  
Therefore, we have extended the Bjorken formula, first by including
the finite time (but neglecting the finite width along the beam
direction $z$) of the initial energy production~\cite{Lin2018}, and 
later by including both the finite time and finite width in
$z$~\cite{Mendenhall2021}.  Figure~\ref{fig1} shows the 
schematic picture in the $z-t$ plane for calculating the energy
density at mid-spacetime-rapidity (inside $-d<z<d$ at finite time $t$
with $d \to 0$) for central A+A collisions. 
The Bjorken formula~\cite{Bjorken1983} assumes that partons are
initially produced at $z_0=0$ and time $0$, 
and the first extension~\cite{Lin2018} 
assumes that partons are initially produced at $z_0=0$ but at any time 
within $[0,d_t]$. In the second extension~\cite{Mendenhall2021}, 
partons can be initially produced anywhere inside the rhombus, which
is within $z_0 \in [-\beta d_t/2, \beta d_t/2]$ and time $\in
[0,d_t]$. Note that parton interactions after their initial productions
are neglected in all three methods.

In the second extension study that considers the full finite thickness
effect~\cite{Mendenhall2021}, the initial energy density
at time $t$ averaged over the full transverse overlap area can be
written as
\begin{equation}
\epsilon(t)=\frac{1}{A_T}\int_S\frac{dxdz_0}{t-x}\frac{d^3m_T}{dxdz_0dy}\cosh^3y.
\label{eden}
\end{equation}
Here, $S$ represents the production area over the initial
production time $x$ and longitudinal position $z_0$ at observation
time $t$, as indicated by the shaded area in Fig.\ref{fig1}. 
For the transverse energy density $d^3m_T/(dxdz_0dy)$, 
we make the simplest assumption that it is uniform over the $z_0-x$
plane, i.e., $d^3m_T/(dxdz_0dy) \propto dm_T/dy$. 
After parameterizing the $dm_T/dy$ function with the measured  
transverse energy rapidity density and net-proton
$dN/dy$~\cite{Mendenhall2021}, we can then calculate the time  
evolution of the energy density. 
We apply the same method to calculate the time evolution of
the net-charge densities~\cite{Mendenhall:2021maf}. 
For example, the net-baryon density from our semi-analytical model is
given by 
\begin{equation}
n_B(t)=\frac{1}{A_T}\int_S\frac{dxdz_0}{t-x}\frac{d^3N_B}{dxdz_0dy}\cosh^2{y}, 
\label{nBden}
\end{equation}
where $N_B$ represents the net-baryon number in an event. 
On the other hand, the net-baryon density in the  Bjorken model is given by
\begin{equation}
n_B^{Bj}(t)=\frac{1}{tA_T}\frac{dN_B}{dy}.
\label{nB_bj}
\end{equation}
Details of the calculations including the parameterizations of 
$dm_T/dy$ and the net-proton $dN/dy$ can be found in the full
studies~\cite{Mendenhall2021,Mendenhall:2021maf}. 
Note that our method enforces the relevant conservation
laws, i.e., the produced matter in a central heavy ion
collision is assumed to have a total energy $A \snn$, total net-baryon
number $2A$,  total net-charge $2Z$, and total net-strangeness 0. 

\section{Results}

\begin{figure}[hbt]
\includegraphics[width=0.8\linewidth]{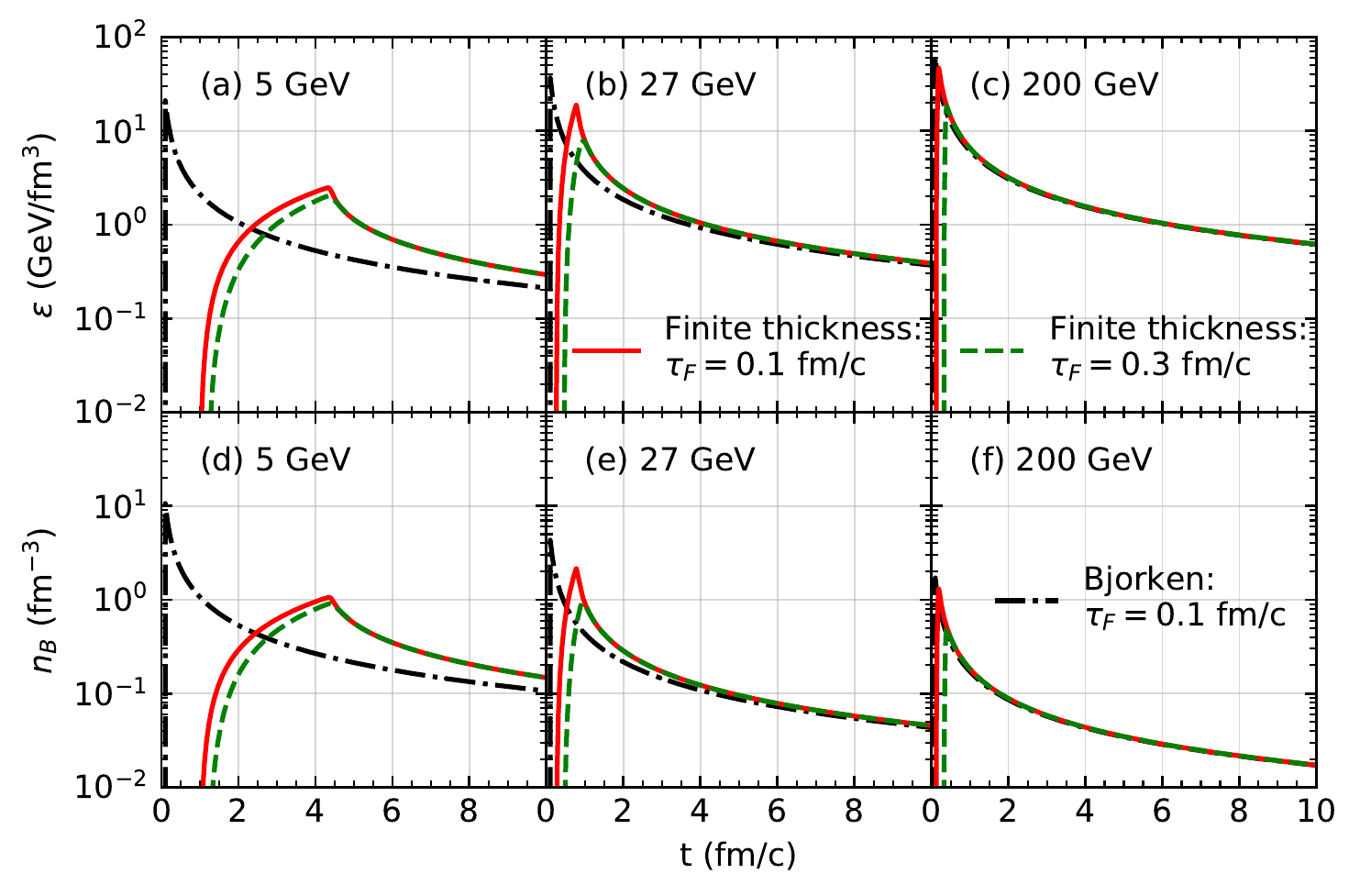}
\caption{(a-c) Energy density $\epsilon(t)$ and (d-f) net-baryon
density $n_B(t)$ at mid-rapidity for central Au+Au
collisions at $\snn=$5, 27 and 200 GeV from our model for
$\tau_F=0.1$ and $0.3$ fm/$c$ in comparison with the Bjorken formula
for $\tau_F=0.1$ fm/$c$.}  
\label{fig2}
\end{figure}

The top panels of Fig.~\ref{fig2} show the results of $\epsilon(t)$,  
the average energy density at mid-rapidity as calculated with
Eq.\eqref{eden}, for central  Au+Au collisions at three different
energies.  The results from our semi-analytical model at $\tau_F=0.1$
(solid) and $0.3$ (dashed) fm/$c$ are shown in comparison with the
results from the Bjorken energy density formula at $\tau_F=0.1$ fm/$c$
(dot-dashed).  The lower panels show the corresponding results of the
net-baryon density $n_B(t)$. 
Compared to the results from the Bjorken formula, our results show 
significantly lower peak values, $\epsilon^{max}$ and $n_B^{max}$, at
lower collision energies, as expected from earlier studies of the
effect of the finite nuclear thickness~\cite{Lin2018,Mendenhall2021}. 
At high collision energies, our results approach the Bjorken formula.
We also see that the peak energy density $\epsilon^{max}$ increases as
$\snn$ increases, while the peak net-baryon density $n_B^{max}$ 
first increases and then decreases with $\snn$. 
Note that for the net-charge and net-strangeness densities, our semi-analytical
method gives the following:
\begin{equation}
n_Q(t)=\frac{Z}{A}n_B(t), ~n_S(t)=0.
\end{equation}
Note that the above relations are often used to constrain the equation of
state such as those from lattice QCD
calculations~\cite{Noronha-Hostler:2019ayj}. 

\begin{figure}[hbt]
 \includegraphics[width=0.8\linewidth]{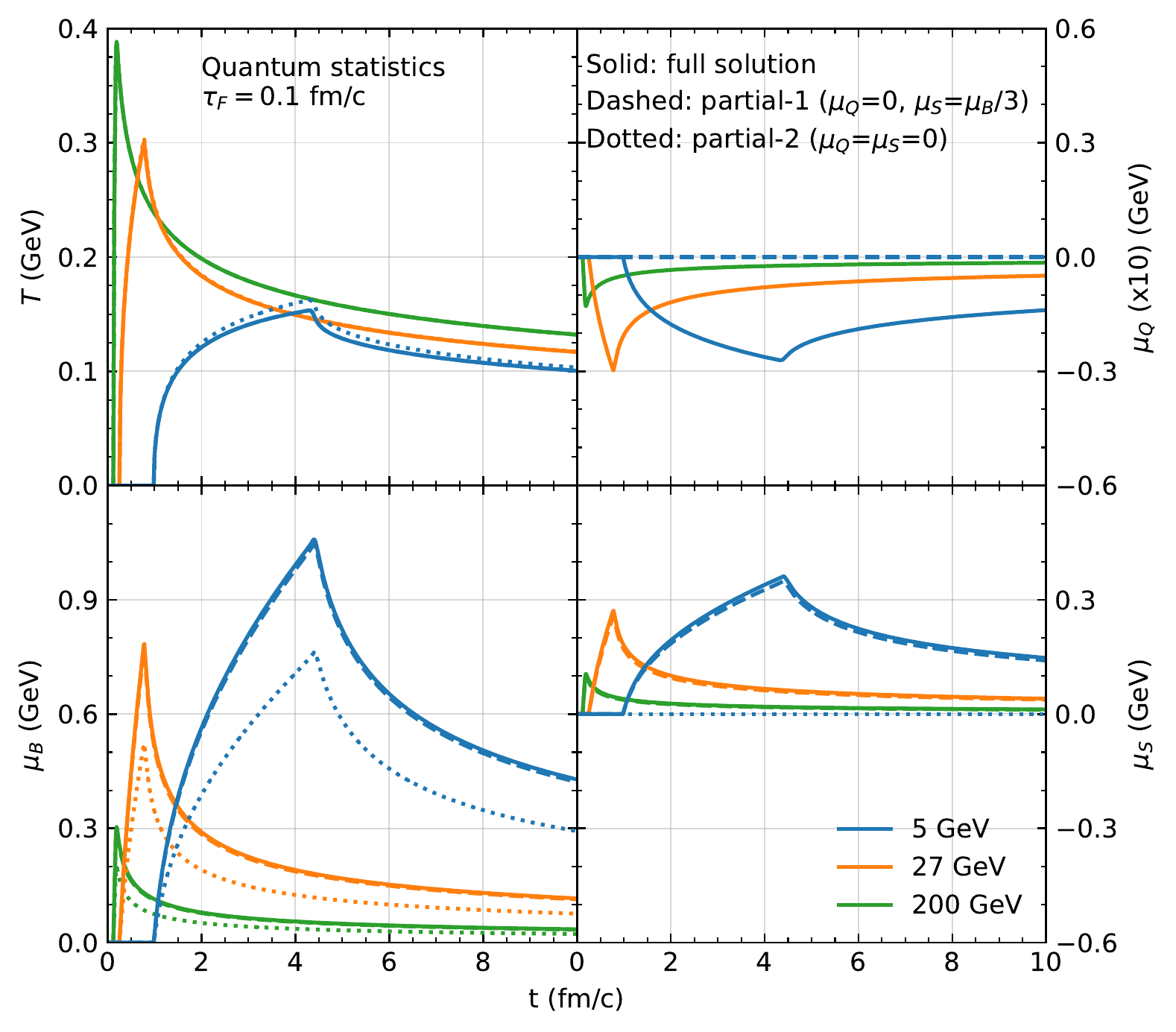}
\caption{The extracted $T(t)$, $\mu_B(t)$, $\mu_Q(t)$, and $\mu_S(t)$
  for $\tauf=0.1$ fm/$c$ from the full solution or partial solutions
  of the quantum ideal gas EOS  for central Au+Au collisions at
  5,   27, and 200 GeV.  Note that the $\mu_Q(t)$ curves have been 
  multiplied by a factor of 10.}
\label{fig3}
\end{figure}

After calculating the densities, we can then use a given equation of
state of the nuclear matter to convert them into the thermodynamical 
quantities: temperature $T$ and chemical potentials $\mu$. 
In this proceeding, we use the ideal gas quark-gluon plasma EOS with
quantum or Boltzmann statistics for the
conversions~\cite{Mendenhall:2021maf}.  The solid curves in
Fig.~\ref{fig3} show the $T$ and $\mu$ results for central Au+Au
collisions at $\tauf=0.1$ fm/$c$, which are extracted from the full
solution of the quantum ideal gas EOS, i.e.,  by solving the four
equations relating   $\epsilon, n_B, n_Q$, and $n_S$ to $T, \mu_B,
\mu_Q$, and $\mu_S$. 
We can see that the peak temperature $T^{max}$ increases with $\snn$,
while the peak net-baryon chemical potential $\mu_B^{max}$ 
decreases with $\snn$. In addition, both $T^{max}$ and $\mu_B^{max}$
are reached earlier at higher energies due to the shorter crossing time.  

If we are only interested in the collision trajectory in the $T-\mu_B$
plane (instead of the four-dimensional $T-\mu$ space) of the QCD phase
diagram, it would be convenient to have the (partial) relations between  
$\epsilon, n_B$ and $T, \mu_B$ without the net-charge or
net-strangeness variables. 
The easiest way to achieve this, which has often been used
~\cite{Noronha-Hostler:2019ayj,Grefa:2021qvt}, is to assume
$\mu_Q=\mu_S=0$. We name the resultant relations as the partial-2
solution of the EOS. However, this assumption violates the strangeness
neutrality condition $n_S=0$ that is expected for heavy ion
collisions.  Indeed, we see from Fig.~\ref{fig3} that the partial-2
solution (dotted curves) gives a much smaller $\mu_B$ than the full
solution and thus cannot give accurate trajectories. 

An alternative way is to assume the following:
\begin{equation}
\mu_Q=0, ~\mu_S=\frac{\mu_B}{3}, 
\end{equation}
and we name the resultant relations as the partial-1 solution of the
EOS. The assumption $\mu_Q=0$ is made because the $\mu_Q$ values 
extracted from the full solution are very small, as shown in the
upper-right panel of Fig.\ref{fig3} where the $\mu_Q$ values have been
multiplied by a factor of 10. 
The smallness of $\mu_Q$ is a consequence of the fact that most nuclei
have $Z \sim A/2$, because  the strangeness
neutrality $n_S=0$ plus the condition $n_Q=n_B/2$ would lead to
$\mu_Q=0$ for either the quantum or Boltzmann ideal gas
EOS~\cite{Mendenhall:2021maf}.  
The other assumption $\mu_S=\mu_B/3$ is made to satisfy the
strangeness neutrality, which requires $\mu_B-\mu_Q-3\mu_S=0$ for the
QGP ideal gas equations of state~\cite{Mendenhall:2021maf}. 
Note that the recent numerical results of the collision trajectories
from the AMPT model~\cite{Wang:2021owa} also show $\mu_Q\approx 0$ and
$\mu_S  \approx \mu_B/3$. 
In the left panels of Fig.\ref{fig3}, we see that the $T(t)$ and
$\mu_B(t)$ results from the partial-1 solution and those from the full
solution agree very well. This demonstrates that the partial-1
solution and its following relations from the quantum ideal gas EOS
are quite accurate, at least for the QGP ideal gas equations of state:
\begin{equation}
\begin{split}
\epsilon_1&=\frac{19\pi^2}{12}T^4+\frac{\mu_B^2}{3}T^2+\frac{\mu_B^4}{54\pi^2},\\
n_{B,1}&=\frac{2\mu_B}{9}T^2+\frac{2\mu_B^3}{81\pi^2}.
\end{split}
\label{partial1}
\end{equation}

\begin{figure}[hbt]
 \includegraphics[width=0.8\linewidth]{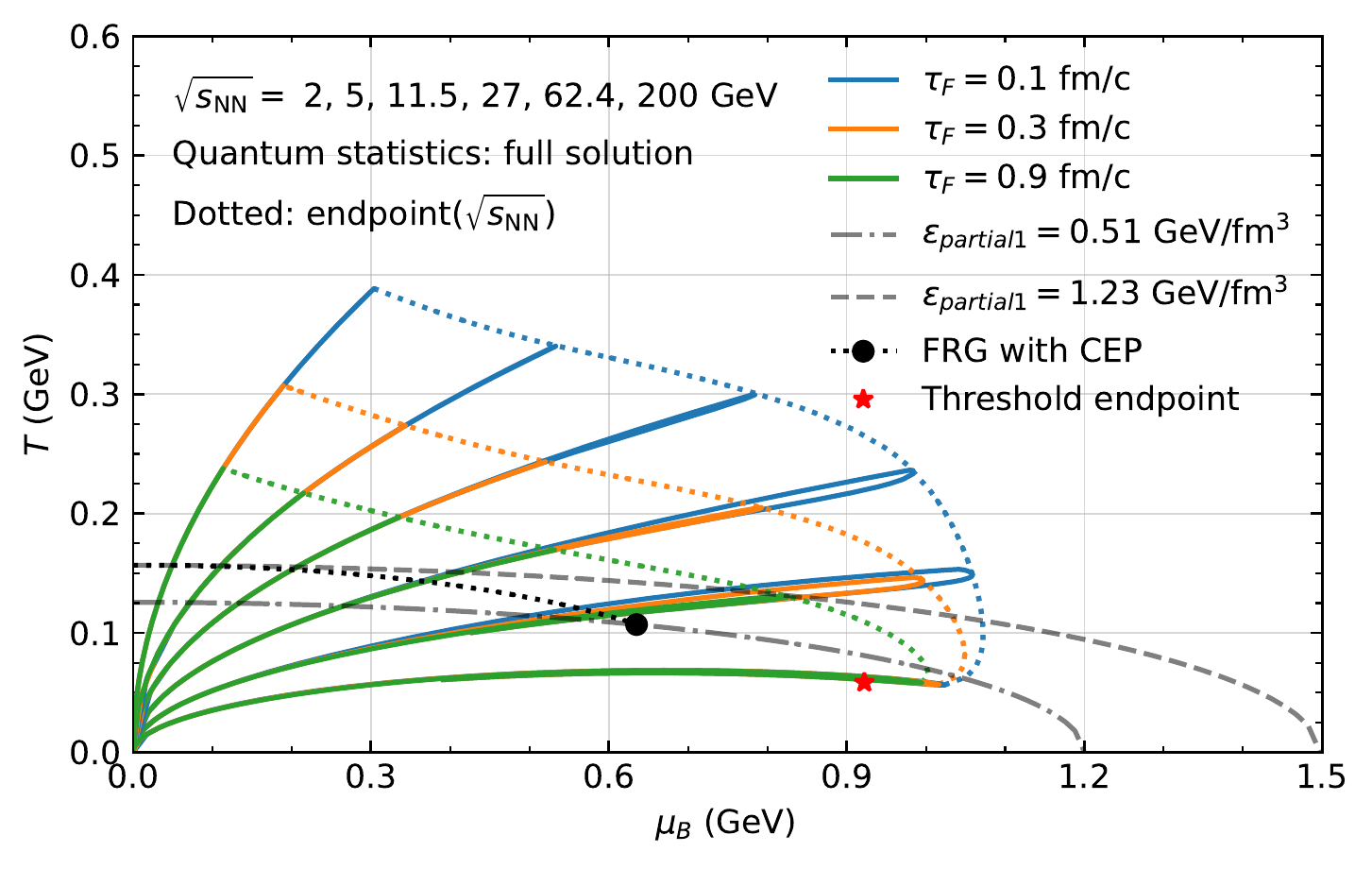}
\caption{Trajectories for the quantum ideal gas EOS at $\tauf=$ 0.1,
  0.3, and 0.9 fm/$c$ for  central Au+Au collisions at different energies, 
together with the trajectory endpoint curves for 
$\snn$ from 2 to 200 GeV. 
The endpoint at the threshold energy is shown as star.
The FRG crossover curve with the critical end point is also shown,
where the two constant energy density curves from the partial-1
solution intersect the endpoints of the FRG curve.}
 \label{fig4}
\end{figure}

Figure~\ref{fig4} shows the $T-\mu_B$ trajectories in the QCD
phase diagram for the quantum ideal gas EOS  at three different
$\tauf$ values for central Au+Au collisions at different energies. 
The QCD crossover curve and the critical end point
at $(\mu_B,T)=(0.635 {\rm ~GeV}, 0.107 {\rm ~GeV})$,  
calculated from the functional renormalization group (FRG)
with $N_F=2+1$~\cite{Fu2020}, are also shown. 
We have also used the quantum partial-1 solution Eq.\eqref{partial1}
to calculate the lines of constant $\epsilon$, which go 
through the two endpoints of the FRG crossover curve at
$\epsilon=1.23$ (dashed) and 0.51 (dot-dashed) GeV/fm$^3$. 
At the threshold energy $\snn=2m_N$ with $m_N$ being the nucleon mass, 
our model gives  $\epsilon^{max}=2\rho_0m_N$ and
$n_B^{max}=2\rho_0$~\cite{Mendenhall:2021maf} ($\rho_0\approx 0.17$
fm$^{-3}$),  which would be expected if the two nuclei would just
fully overlap.  If we treat this matter as an ideal gas QGP  with
quantum statistics, it will be located at $(\mu_B,T) \sim (0.9
{\rm~GeV}, 0.06 {\rm ~GeV})$, as shown by the star symbol in
Fig.~\ref{fig4}.  When a trajectory reaches the endpoint, where both
$\epsilon^{max}$ and $n_B^{max}$ are reached, it turns clockwise and
returns toward the origin. We see that the returning part of the
trajectory  almost overlaps with the outgoing part at high collision
energies.  In addition, the trajectories  in Fig.~\ref{fig4} pass
through the crossover  curve for central Au+Au collisions at $\snn
\gtrsim 4$ GeV. 

The results from our semi-analytical model
~\cite{Lin2018,Mendenhall2021,Mendenhall:2021maf} depend on the value
of $\tauf$, and Fig.~\ref{fig4} also shows how the  $T-\mu_B$
trajectories  from the quantum ideal gas EOS depend on $\tauf$. 
At a smaller $\tauf$, the peak densities are higher; therefore, the
trajectory becomes longer with the endpoint moving further to higher
$\mu_B$ (and also higher $T$ except at very low energies). 
The endpoints of the trajectories as functions of the collision energy 
are shown by the three colored dotted curves in Fig.~\ref{fig4} 
for three different formation times. 
We observe a clear separation of the endpoint curves of different
$\tauf$ values, except for very low collision energies, 
where the endpoint curves become less sensitive to $\tauf$. 
This is expected because the densities such as the energy density
depend on  $\tauf$ weakly at low collision energies but strongly 
($\propto 1/\tauf$) at high energies~\cite{Mendenhall2021}. 
We also see in Fig.~\ref{fig4} that even for the relatively large
value of $\tauf=0.9$ fm/$c$,  the CEP from the FRG calculation is
within the coverage of the trajectory endpoint curve. 

\begin{figure}[hbt]
 \includegraphics[width=0.8\linewidth]{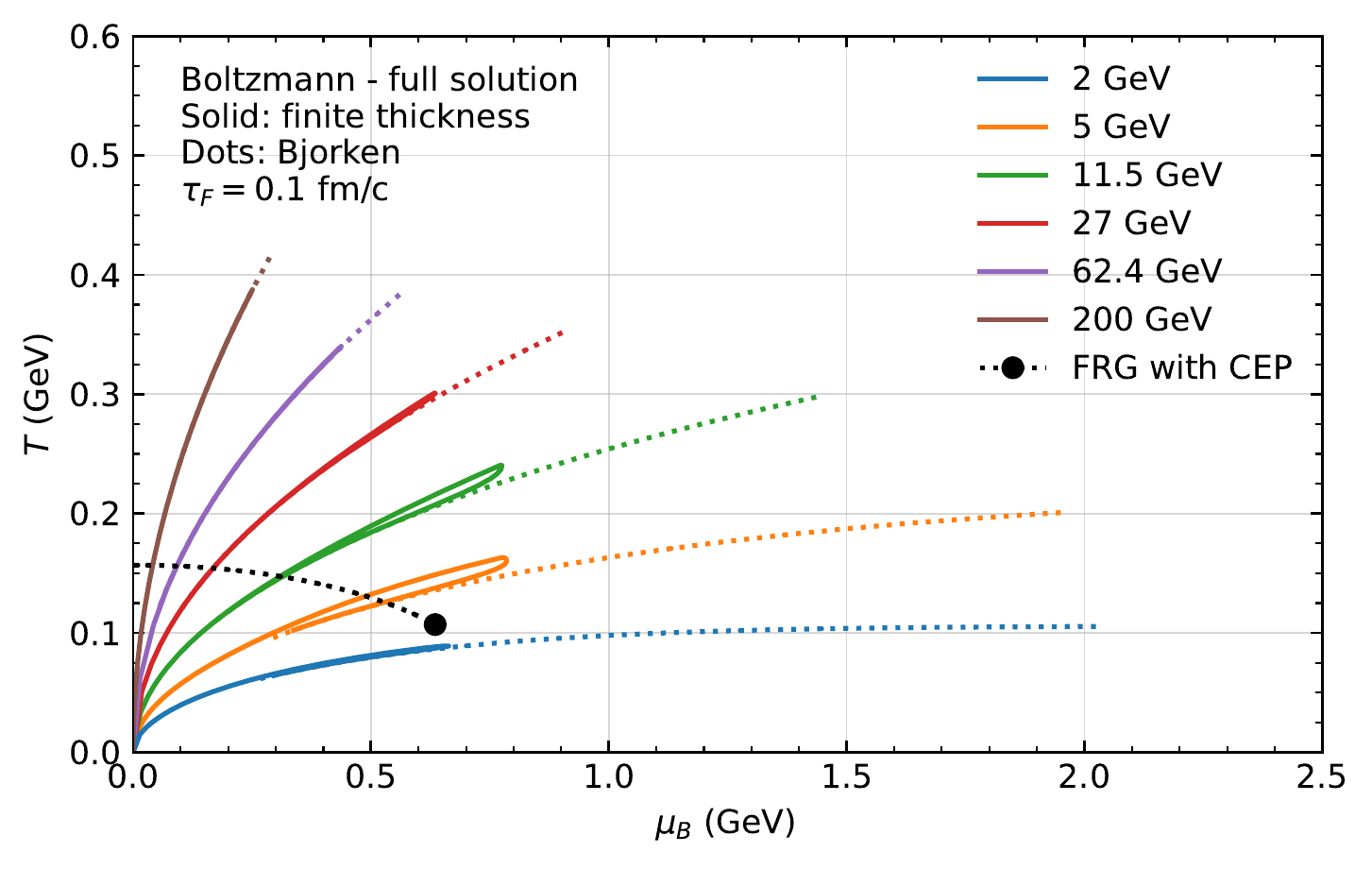}
\caption{Trajectories for the Boltzmann ideal gas EOS
for central Au+Au collisions at different
energies from our method and the Bjorken
formula for $\tauf=0.1$ fm/$c$. The FRG crossover curve with the CEP
is also shown.} 
 \label{fig5}
\end{figure}

To see the effect of the finite nuclear thickness, in
Fig.~\ref{fig5} we compare our trajectories with the Bjorken
trajectories, i.e., trajectories extracted from the $\epsilon(t)$ and
$n_B(t)$  values calculated with the Bjorken formula, for the
Boltzmann ideal gas EOS. At high collision energies, our trajectories
are rather close to the Bjorken trajectories as expected.  At lower
collision energies, however, the $\mu_B^{max}$ value from the 
Bjorken trajectory is much larger because of its much higher peak
density $n_B^{max}$. 
In addition, we see in Fig.~\ref{fig5} that the late-time part of the 
Bjorken trajectory overlaps with the returning part of our
trajectory.  This is expected because at late times our
semi-analytical model  approaches the Bjorken
formula~\cite{Lin2018,Mendenhall2021}, which can be seen in
Fig.~\ref{fig2}.

By comparing the trajectories from the quantum ideal gas EOS in
Fig.~\ref{fig4} and those from the Boltzmann ideal gas EOS in
Fig.~\ref{fig5}, we can see that the trajectory depends on the
equation of state. While the $T^{max}$ values are often similar at the
same collision energy (except for very low energies), the
$\mu_B^{max}$ value is significantly larger in the quantum ideal gas
EOS. This feature is also seen in the trajectories calculated from the
AMPT  model~\cite{Wang:2021owa} and can be understood in terms of the
Pauli exclusion principle in the quantum EOS. 

\section{Summary and outlook}

We have developed a semi-analytical model to calculate the
time-dependent  energy density $\epsilon(t)$, net-baryon density
$n_B(t)$, net-electric-charge density $n_Q(t)$, and net-strangeness
density $n_S(t)$ at mid-pseudorapidity averaged over the transverse
overlap area in central Au+Au collisions.   
We then extract the time evolution of the thermodynamical quantities
$T, \mu_B, \mu_Q$, and $\mu_S$ assuming the
formation of a QGP with either quantum or Boltzmann ideal gas equation
of state. This enables us to plot the collision trajectories in the
$T-\mu_B$ plane of the QCD phase diagram. 

The trajectories from our model are very different from those
calculated with the Bjorken formula at energies below tens of GeVs, 
demonstrating the importance of including the finite nuclear thickness
at those energies. 
We also find that the accessible area in the phase diagram 
depends strongly on the parton formation time $\tau_F$ when the 
collision energy is higher than a few GeVs. 
However, even when using a relatively large $\tau_F$ value of 0.9
fm$/c$, the critical end point from the FRG calculation is within the area
covered by the trajectories. 
We also find that the $T-\mu_B$ results from the simpler partial
solution that assumes $\mu_Q=0$ and $\mu_S=\mu_B/3$ are very close to
the full solution. On the other hand, the results from another partial
solution that assumes $\mu_Q=\mu_S=0$, which violates the strangeness
neutrality, significantly underestimate  the extracted $\mu_B$ values. 

We find that the collision trajectory depends on the equation of
state. For our results from the ideal gas equations of state for the
quark-gluon plasma, the calculated trajectory should break down 
soon after it goes below the crossover curve (or the first-order phase
transition curve) in the QCD phase diagram. 
We plan to extend this study by using more realistic equations of
state, such as those based on the lattice QCD
calculations~\cite{Noronha-Hostler:2019ayj}.
In addition, we have so far neglected the transverse expansion of the
created matter, which would decrease  the peak densities and thus
affect the collision trajectory in the phase  diagram. We also plan to
include this effect in the update of the full
study~\cite{Mendenhall:2021maf}. 

We have written a web interface~\cite{interface}, which will plot the
calculated energy density as a function of time as well as the event
trajectory in the T-$\mu_B$ plane according to the user's input for
the colliding system, energy and formation time $\tauf$. A data file
for the time evolution of the energy density, temperature, and the
three chemical potentials can also be downloaded. So far only the
ideal gas equations of state are implemented at the web interface, and
we plan to add a more realistic equation of state. 
We hope that this semi-analytical model will provide the community a
useful tool for  exploring the trajectories of nuclear collisions in
the QCD phase diagram in the $T-\mu_B$ plane or the general
$T-\mu_B-\mu_Q-\mu_S$ space.

\section*{Acknowledgments}
This work has been supported by the National Science
Foundation under Grant No. PHY-2012947.

\section*{References}
\bibliography{ref}

\end{document}